\documentclass[preprint,proceedings]{rmaa}
\usepackage{paralist}

\usepackage{psfrag,color}

\SetYear{2002}
\SetConfTitle{Galaxy Evolution: Theory and Observations}

\title{New Tests of the Cluster Entropy Floor Hypothesis} 

\author{
  I. G. McCarthy,\altaffilmark{1}
  A. Babul,\altaffilmark{1}
  M. L. Balogh,\altaffilmark{2}
  and G. P. Holder\altaffilmark{3}}

\altaffiltext{1}{University of Victoria, Victoria, BC, Canada.}
\altaffiltext{2}{University of Durham, Durham, UK.}
\altaffiltext{3}{Institute for Advanced Study, Princeton, NJ, USA.}

\shortauthor{McCarthy et al.}
\shorttitle{Excess Entropy in the ICM}

\fulladdresses{
\item Arif Babul and Ian G. McCarthy: Dept. of Physics and Astronomy, 
University of Victoria, Victoria, BC, V8W 3P6, Canada (babul, 
imccarth@uvic.ca).
\item Michael L. Balogh: Dept. of Physics, University of Durham,
South Road, Durham, DH1 3LE, UK (M.L.Balogh@durham.ac.uk).
\item Gilbert P. Holder: School of Natural Sciences, Institute for Advanced Study,
Princeton, NF, 08540, USA (holder@ias.edu).}

\listofauthors{I. G. McCarthy, A. Babul, M. L. Balogh, \& G. P. Holder}
\indexauthor{McCarthy, I. G.}
\indexauthor{Babul, A.}
\indexauthor{Balogh, M. L.}
\indexauthor{Holder, G. P.}

\abstract{Recent efforts to account for the observed $L_X - T_X$ relation of galaxy 
clusters has led to suggestions that the ICM has an apparent ``entropy floor'' at the 
level of $K_0 \gtrsim 300$ keV cm$^2$.  Here, we propose new tests based on the thermal 
SZ effect and on the $M_{gas} - T_X$ trend (from X-ray data) to probe the level of the 
excess entropy in the ICM.  We show that these new tests lend further support to the 
case for a high entropy floor in massive clusters.}

\addkeyword{Cosmology: Theory}
\addkeyword{X-rays: Galaxies: Clusters}

\begin{document}
% Typeset article header
\maketitle

\section{Introduction}
\label{sec:intro}

Relationships between the global X-ray properties of clusters have proven to be important 
probes of the intracluster medium (ICM).  Case in point are studies of the X-ray 
luminosity ($L_X$) - mean emission-weighted gas temperature ($T_X$) relation.  
Theoretical models that include only the effects of gravity and shock heating 
(self-similar models) predict $L_X \propto T_X^2$, yet the observed relation is $L_X 
\propto T_X^{2.6-3.0}$ (e.g., Markevitch 1998; Allen \& Fabian 1998).  This discrepancy 
has prompted a number of theorists to consider alternative models.  Both the effects of 
heating (e.g., Kaiser 1991; Evrard \& Henry; Wu et al.\@ 2000; Babul et al.\@ 2002) and 
radiative cooling (e.g., Bryan 2000; Voit \& Bryan 2001; Dav\'{e} et al.\@ 2002) have 
been examined.  These studies find that heating and/or cooling introduces a core into 
the entropy profiles of clusters (an ``entropy floor'') which, in turn, results in a 
steepening of the $L_X - T_X$ relation (as required).  Ponman and collaborators (Ponman 
et al.\@ 1999; Lloyd-Davies et al.\@ 2000) have presented direct evidence for an entropy 
floor in nearby groups.

Investigations of the ICM are not limited to the $L_X - T_X$ relation, however.  Other 
cluster observables can be used as alternative probes of the ICM.  For example, McCarthy 
et al.\@ (2002) have studied the effects of an entropy floor on the cluster gas mass 
($M_{gas}$) - $T_X$ relation and, through a detailed comparison with observations, 
have placed stringent limits on the entropy floors in nearby massive clusters.  Because 
the $M_{gas} - T_X$ relation is derived from X-ray data, the test provides a valuable 
self-consistency check of the $L_X - T_X$ results.  The main results of that study are 
presented below.

Ultimately, however, scaling relations that are independent of the $L_X - T_X$ and 
$M_{gas} - T_X$ relations are desirable.  The thermal Sunyaev-Zeldovich (SZ) 
effect, which has a different dependence on the entropy of the ICM than does the X-ray 
emission, can be used for such a purpose (McCarthy et al.\@ in preparation).  Here, we 
derive a relation between the central and integrated cluster Compton parameters (which 
are both proportional to the SZ effect), analyze how this relation is affected by an 
entropy floor, and compare the results to recent SZ effect observations.  This is the 
first time the SZ effect has been used as a probe of the entropy floors of clusters. 

\begin{figure}[!t]
\includegraphics[width=\columnwidth]{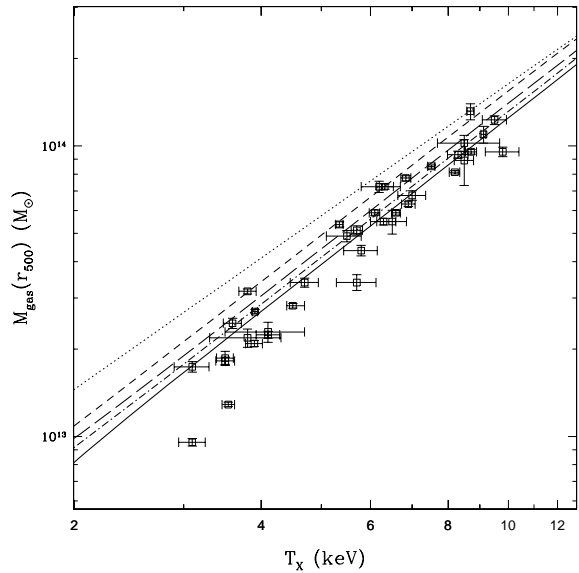}
\caption{A comparison of $M_{gas}(r_{500}) - T_X$ relations.  The squares represent the
observations of Mohr et al. (1999).  The dotted line is the self-similar result.
The short-dashed, long-dashed, dot-dashed, and solid lines represent the models with
entropy floor constants of $K_0$ = 100, 200, 300, and 427 keV cm$^2$, respectively.}
\label{fig1}
\end{figure}

\section{The $M_{gas} - T_X$ Relation}

In Figure 1 we present the $M_{gas} - T_X$ relation as predicted by the
Babul et al.\@ (2002) analytic models within $r_{500}$, the radius within which the
mean dark matter mass density of the cluster is 500 times the mean critical density at 
$z$ = 0.  When compared to the observational data of Mohr et al.\@ (1999), we find that 
only the models with a high entropy floor ($K_0 \gtrsim 300$ keV cm$^2$) are consistent 
with the data.  The self-similar model is ruled out ($>$ 99\% confidence).  

\section{A SZ effect Scaling Relation}

Figure 2 is a plot of the predicted $y_0 - S_{\nu}(r < 150$ kpc)$/f_{\nu}$ relations, 
where $y_0$ is the Compton parameter evaluated through the cluster center and $S_{\nu}(r 
< 150 $kpc$)/f_{\nu}$ is proportional to the integrated Compton parameter within the 
central (projected) 150 kpc.  The lines hold the same meanings as in Figure 1.  Again, we 
find that only models with $K_0 \gtrsim 300$ keV cm$^2$ are able to match the SZ 
effect observations of Reese et al.\@ (2002; McCarthy et al.\@ in preparation). 

\section{Summary}

\begin{figure}[!t]
\includegraphics[width=\columnwidth]{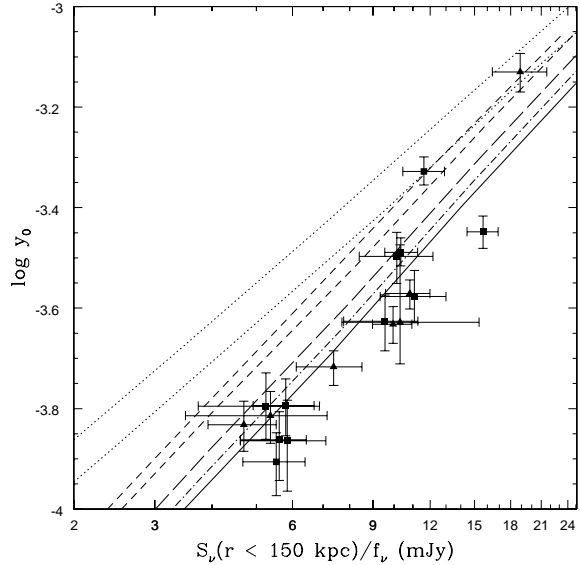}
\caption{A comparison of $y_0 - S_{\nu}(r < 150 $kpc$)/f_{\nu}$ relations.  The
squares ($0.14 \leq z \leq 0.3$) and triangles ($z > 0.3$) represent the data of Reese
et al. (2002).  The thick, thin lines are the $z$ = 0.2, 0.5 predictions,
respectively.  For clarity, we plot the $z = 0.5$ lines for the self-similar and $K_0
= 100$ keV cm$^2$ models only.  The integrated Compton parameters (both data and
models) have been arbitrarily rescaled for $z = 0.2$.}
\label{fig2}
\end{figure}

The relations that we have described above demonstrate that a high entropy 
floor ($K_0 \gtrsim 300$ keV cm$^2$) is required to match the X-ray and SZ
effect observations of massive clusters.  This is consistent with 
previous investigations of the $L_X - T_X$ relation (e.g., Tozzi \& 
Norman 2001).  More work
\adjustfinalcols is required to determine the origin of the entropy floor.

\end{document}